\documentclass[epjST]{svjour}
\usepackage{graphics,csquotes}
\usepackage[english]{babel}
\usepackage{verbatim,bbm,mathrsfs,amsmath,amssymb}
\usepackage{graphicx,color,epsfig,enumerate}
\newcommand{\ket}[1]{\vert {#1} \rangle}
\newcommand{\bra}[1]{\langle{#1}\vert}
\newcommand{\ketbra}[2]{\vert {#1} \rangle \langle{#2}\vert}

\begin{document}
\title{Photon pair production by STIRAP in ultrastrongly coupled matter-radiation systems}
\author{A. Ridolfo\inst{1}  \and G. Falci \inst{2,3,4}\fnmsep\thanks{\email{gfalci@dmfci.unict.it}} \and F.M.D. Pellegrino \inst{2,4} \and G.D. Maccarrone \inst{2}
\and E. Paladino \inst{2,3,4}}
\institute{RIKEN, Saitama 351-0198, Japan \and
Dipartimento di Fisica e Astronomia, Universit\`a di Catania, Via S. Sofia 64, Catania (I). \and 
CNR-IMM Catania Universit\'a (MATIS), Via Santa Sofia 64, Catania (I).
 \and Istituto Nazionale di Fisica Nucleare, Via Santa Sofia 64, Catania (I)
}
\abstract{Artificial atoms (AA) offer the possibility to design physical systems implementing new regimes of ultrastrong coupling (USC) between radiation and matter~\cite{ka:205-ciuti-prb-intersubbandpolariton}, where previously unexplored non perturbative physics emerges.
While experiments so far provided only spectroscopic evidence of USC, 
we propose the dynamical detection of virtual photon pairs in the dressed eigenstates, which is a \enquote{smoking gun} of the very existence of USC in nature. We show how to coherently amplify this channel to reach 100\% efficiency by operating advanced control similar to stimulated Raman adiabatic passage (STIRAP)~\cite{kr:217-vitanovbergmann-rmp}.
}
%
\maketitle
\section{Introduction}
\label{sec:intro}
Since 1983 new fundamental phenomena have been demonstrated in systems of two-level atoms coupled to a high-$Q$ cavity~\cite{ka:189-harochekleppner-phystoday}. In the simplest cases the strong coupling (SC) regime is described by the quantum Rabi model~\cite{ka:137-rabi-pr-rabimodel,ka:211-braak-prl-rabisol}
\begin{equation}
H_R = \varepsilon \, \ketbra{e}{e} + 
\omega_c \,a^\dagger a + g \,\big[a \ketbra{e}{g}+ a^\dagger \ketbra{g}{e}\big] +
g \, \big[a^\dagger \ketbra{e}{g}+ a \ketbra{g}{e}\big].
\label{eq:H-rabi-model}
\end{equation}
Here, $\{\ket{g},\ket{e}\}$ are the eigenstates of the two-level atom, $\epsilon$ being their energy splitting , 
and the energy of a quantum in the cavity is $\omega_c \sim \epsilon$. In the SC regime the coupling constant $g$ between the atomic dipole and the quantized electric field, $\propto a^\dagger + a$ is larger than both the decay rate of the atom, $\gamma$, and that of the cavity, $\kappa$. In the last decade this model has been implemented in architectures of AAs, such as quantum dots and superconducting Josephson \enquote{circuit-QED} systems~\cite{ka:204-wallraff-superqubit,ka:203-plastinafalci-prb,kr:217-nori-review-supercqed}. 
In the SC regime $g/\omega_c$ is small, typically $\sim 10^{-6}-10^{-2}$ depending on the implementations, and the \enquote{counterrotating} term in the Hamiltonian Eq.~(\ref{eq:H-rabi-model}) can be neglected, yielding the Jaynes-Cummings (JC) model~\cite{kr:217-nori-review-supercqed}. In this limit the dynamics is described in terms of single excitations coherently exchanged between atom and cavity, thus making circuit-QED systems fundamental building blocks for the design of quantum architectures~\cite{kb:210-nielsenchuang}.

\begin{figure}[t!]
\centering
\resizebox{!}{120pt}{
\includegraphics{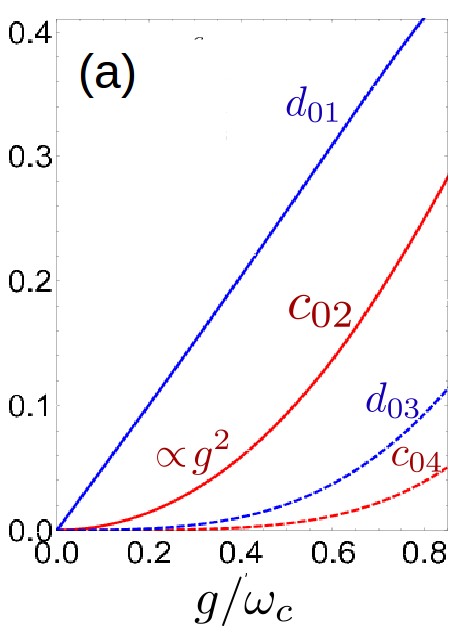}} \quad
\resizebox{!}{120pt}{\includegraphics{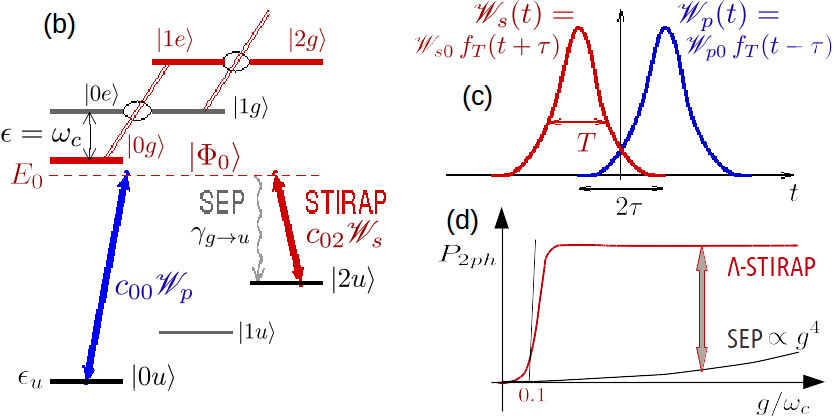}} 
\caption{(a) 
Amplitudes $c_{02}$ and $d_{1\pm,2}$ which are nonvanishing 
for eigenstates $\ket{\Phi_0}$ and $\ket{\Phi_{1\pm}}$ in the USC regime. 
(b) Scheme of a three-level system coupled to an oscillator in the USC regime: population is pumped from $\ket{0\,u}$ to the dressed ground state $\ket{\Phi_0}$, virtual photons pairs being converted to real photons by spontaneous emission or by STIRAP-induced coherent population transfer
to $\ket{2\,u}$. The pump pulse is resonant with the $\ket{0u} \leftrightarrow \ket{\Phi_0}$ transition, $\omega_p = E_0-\epsilon_u$, and the Stokes pulse with the $\ket{\Phi_0} \leftrightarrow \ket{2u}$ transition, $\omega_s =  E_0 - (\epsilon_u + 2 \omega_c)$. (c) STIRAP is operated by shining the Stokes pulse (red) before the pump pulse (blue); we used a Gaussian $f_T(t)$. (d) STIRAP amplifies coherently the photon pair production channel, up to $100\%$ efficiency.}
\label{fig:lambdascheme}       
\end{figure}

AAs offer the possibility to enter the ultrastrong coupling (USC) regime, $g \sim \omega_c, \epsilon$~\cite{ka:205-ciuti-prb-intersubbandpolariton,ka:209-anapparabeltram-prb-ustr,ka:210-niemczyck-natphys-ultrastrong,ka:212-scalari-science-USCTHz,kr:217-nori-review-supercqed,ka:217-yoshiarasemba-natphys-dscflux,ka:216-pellegrino-natcomms,ka:214-pellegrino-prb}
where several new features appear. For instance while the JC ground state is factorized, $\ket{n=0} \otimes \ket{g}$, in the USC regime it contains virtual photons, $\ket{\Phi_0} = \sum_{n=0}^\infty c_{0\,n} \ket{ng} + d_{0n} \ket{n\,e}$ with vanishing $c_{0n}$ for odd number of 
photons $n$ and vanishing $d_{0n}$ for even $n$ (see Fig.~\ref{fig:lambdascheme}a). 
Dynamics in the USC regime is expected to display several new phenomena~\cite{ka:205-ciuti-prb-intersubbandpolariton,ka:207-deliberato-prl-uscdynamics,ka:212-ridolfo-prl-photonblock,ka:217-garziano-acs-phbunching,kr:217-nori-review-supercqed}, which  
however have not been observed so far, experiments being limited to spectroscopy. Therefore demonstrating dynamics in the USC regime remains an outstanding challenge. In this work we will analize an advanced control protocol allowing to detect dynamcally virtual photons in the ground state~\cite{ka:217-falci-fortphys-fqmt,ka:218-falci-usc}. Implementing such a protocol would be an important proof of principle of controlled dynamics in the USC regime. 

\section{Dynamical detection of USC} 
It has been proposed to detect ground-state photons by converting them into real ones letting them decay to an ancillary uncoupled atomic level $\ket{u}$~\cite{ka:213-stassisavasta-prl-USCSEP}, a channel active if and only if $c_{02}\neq 0$,  marking thus USC. The Hamiltonian of this system is $H = \epsilon_u \ketbra{u}{u} + \omega_c \,a^\dagger a\otimes \ketbra{u}{u} + H_R$. For moderate $g/\omega_c$ the above pair production channel has a low yield, as it depends on $|c_{02}|^2 \propto g^4$~\cite{ka:218-falci-usc}. 
We can amplify this channel by driving the system coherently with a two-tone field~\cite{ka:217-falci-fortphys-fqmt},
	$W(t) = \mathscr{W}_s(t) \,\cos (\omega_s t) + \mathscr{W}_p(t) \,\cos (\omega_p t)$ acting on the atom,
	$H_c(t) = W(t) \,\big[ \ketbra{u}{g} +\ketbra{g}{u} \big]$, which 
	implements a $\Lambda$ configuration (see Fig.~\ref{fig:lambdascheme}b)
	 for suitable choice of the frequencies $\omega_k$. 
For not too large $g \lesssim \omega_c= \epsilon$, the dynamics is captured by a simplified three-level model~\cite{ka:217-falci-fortphys-fqmt} described
by the following Hamiltonian in the rotating frame
\begin{equation}
\label{eq:lambda-3}
H_3(t)  = \Big[
\frac{\mathscr{W}_p(t)}{2}\,  \ketbra{0u}{\Phi_0}+
\frac{c_{02}(g) \mathscr{W}_s(t)}{2}\, \ketbra{2u}{\Phi_0}
\Big] + 
\mbox{h.c.}
\end{equation}
Shining the two pulses in the \enquote{counterintuitive} sequence (the Stokes before the pump, Fig.~\ref{fig:lambdascheme}c), this Hamiltonian implements STIRAP~\cite{kr:217-vitanovbergmann-rmp} which yields complete and faithful population transfer $\ket{0u} \to \ket{2u}$ via the intermediate virtual state $\ket{\Phi_0}$. The necessary conditions are that $c_{02}(g) \neq 0$ and that $\max[\mathscr{W}_k(t)]$ are large enough to guarantee adiabatic dynamics, $\sim 100\%$ efficiency being attainable for $g/\omega_c \gtrsim 0.2$ in typical devices (see Fig.~\ref{fig:lambdascheme}d). 
The implementation of $\Lambda$-STIRAP in superconducting AAs has been proposed in the last decade~\cite{ka:206-siebrafalci-optcomm-stirap,ka:208-weinori-prl-stirapqcomp,ka:209-siebrafalci-prb,kr:211-younori-nature-multilevel}, demonstrated only recently~\cite{ka:216-kumarparaoanu-natcomm-stirap,ka:216-xuhanzhao-natcomm-ladderstirap}, and may be a promising tool for processing in complex architectures~\cite{ka:216-vepsalainen-photonics-squtrit,ka:215-distefano-prb-cstirap}. 
In our case detection of a two-photon state in the cavity is a \enquote{smoking gun} for USC, STIRAP providing coherent amplification 
of the USC pair production channel.

\begin{figure}[t]
\centering
\resizebox{!}{110pt}{%
\includegraphics{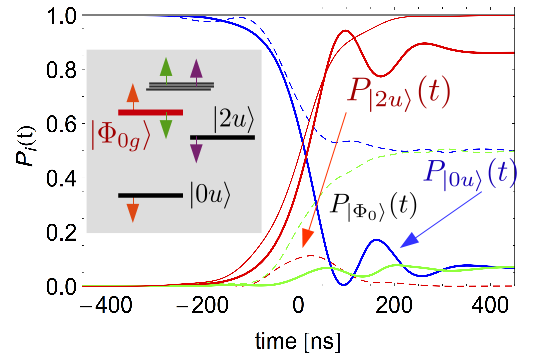}}\quad
\resizebox{!}{110pt}{\includegraphics{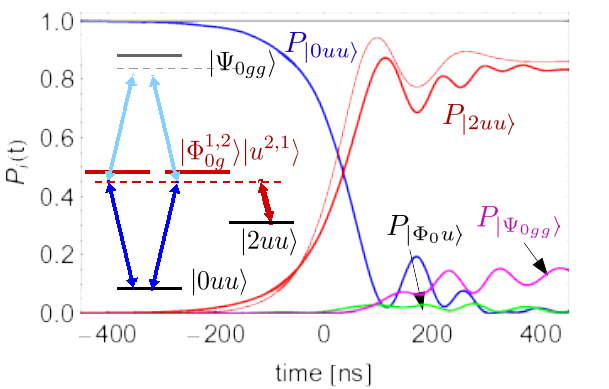}}
\caption{Population histories for systems up to 40 levels, with pulses 
coupled to al the transitions. Paramenters:  
$\epsilon=\omega_c$ and $\epsilon^\prime=4 \omega_c$; pulse 
amplitudes and time scales are suited for superconducting qubits~\cite{ka:210-niemczyck-natphys-ultrastrong,ka:217-yoshiarasemba-natphys-dscflux,ka:218-falci-usc}.
(a) Sinlge atom with $g/\omega_c=0.2$. Dynamical Stark shifts suppressing population transfer in a three-level system (dashed line), are 
partially self-compensated in a multilevel structure (full thick lines) 
and can be eliminated by properly crafted pulses (thin line). 
Inset: Stokes-induced Stark shifts of different sign tend to cancel in a multilevel structures.
(b) Coherent population transfer for two-atoms Rabi model with 
$g=0.2/\sqrt{2}$ (thick lines) is qualitatively equivalent to population transfer for a single atom with coupling $g=0.2$ (thin line), yielding 
already $\gtrsim 90\%$ efficiency with no phase control. In the inset: level scheme of the effective Hamiltonian $H_5$; the pump pulse may address unwanted transitions to the dressed $\ket{0gg}$
state.}
\label{fig:stark}       
\end{figure}
\subsection{Advanced dynamics in multilevel systems}
\label{sec:multilevel}
The simple Hamiltonian Eq.~(\ref{eq:lambda-3}) is valid for small 
enough field amplitudes $\mathscr{W}_k$. This is not the case
if $g$ is small since attaining adiabaticity would require a large $\mathscr{W}_s$. The main effect is a dynamical Stark shift 
of the $\ket{0u} \leftrightarrow \ket{\Phi_0}$ splitting, which   
translates in stray nonvanishing detunings during the protocol. 
In particular the two-photon detuning $\delta(t)=\Delta(E_0 - \epsilon_{u})/2$ is comparable to $ c_{02} \mathscr{W}_s$, and may suppresses 
STIRAP (dashed curves in Fig.~\ref{fig:stark}a). 

Actually the three-level analysis must be generalized to account for the 
multilevel nature of the system~\cite{ka:218-falci-usc}. Fortunately this turns out to mitigate the detrimental effect of dynamical Stark shifts. We studied this problem by considering up to 40 levels and a control field with the complete structure $H_c(t)= W(t) [(\ketbra{u}{g} + (1/\eta) \ketbra{g}{e}) + \mbox{h.c.} ]$, which describes the experimentally relevant case of a ladder type \enquote{dipole} coupling to the AA,  
$\eta$ being the ratio between the corresponding matrix elements. Results in  Fig.~\ref{fig:stark}a (red curves)
show that experimentally detectable population transfer is achieved also in this non-ideal case.
This is due to the fact that the large Stokes field couples also to transitions between higher energy levels, and produces dynamical Stark 
shifts with the same time dependence of the stray $\delta(t)$, which 
it may be partially compensated (see the inset of Fig.~\ref{fig:stark}a).

Moreover, the full signal can be recovered if appropriately crafted 
control is used (thin red curve in Fig.~\ref{fig:stark}a). One option is to use a phase modulation of the Stokes pulse, as explained in~\cite{ka:216-distefano-pra-twoplusone},
designed to compensate the effect of the Stokes field coupled to the $u\!-\!g$ transition only. The fact that this is the relevant source of noise is suggested by success of the strategy even when the 
drive couples to all the ladder transitions of the AA. 
A simpler option is to add a suitable off-resonant tone 
$W(t) \to W(t) + \mathscr{W}_s(t) \,\cos (2 \omega_s t)$, 
whose design prescriptions will be discussed elsewhere~\cite{kn:fischio}.

\section{Amplification by many atoms}
Using an ensemble of $N$ atoms is expected to amplify interaction effects, 
yielding an effective $g \to \sqrt{N} g$, at least for very weakly coupled atoms. The spectrum of the $N$ atoms Rabi Hamiltonian shows this equivalence even for nonperturbative values of $g$. 
This is the mechanism for obtaining USC in semiconductors subband polaritons~\cite{ka:209-anapparabeltram-prb-ustr}, behavng as a system with very large
$N$, but it is also interesting to exploit the limit of few more strongly coupled atoms, which is likely to  
be implemented in the near future in superconducting architectures~\cite{kr:217-nori-review-supercqed}. 
In this limit STIRAP is operated by the control Hamiltonian 
$H_c = W(t) \sum_{k=1}^N (\ket{u^k} \bra{g^k} + \mbox{h.c.})$. 
Fig.~\ref{fig:stark}b shows the protocol for two identical atoms coupled to a mode with a rather small $g/\omega_c=0.2/\sqrt{2}$ compared 
with the result for a single atom with $g=0.2$, showing the expected qualitative agreement. To illustrate the main features we analize the simple situation of resonant and not too strong drives. In this limit the relevant dynamics reduces to five levels (see inset in Fig.~\ref{fig:stark}b) with Hamiltonian 
\begin{equation}
\label{eq:lambda-5}
\begin{aligned}
H_5(t)  &= 
\Big\{ 
\Big(\frac{\mathscr{W}_p(t)}{2}\,\ket{0u u} 
+ 
\frac{c_{02}(g) \mathscr{W}_s(t)}{2}\, 
\ket{2uu} \Big) \,\big[
\bra{\Phi_0^1} \bra{u^2}+
\bra{\Phi_0^2} \bra{u^1} \big]
\\
&+ \frac{\mathscr{W}_p(t)}{2}\,
\big[\ket{\Phi_0^1} \ket{u^2}+
\ket{\Phi_0^2} \ket{u^1} \big] \bra{\Psi_{0gg}}
\Big\} + 
\mbox{h.c.} + (E_{0gg}- 2 E_0) \ketbra{\Psi_{0gg}}{\Psi_{0gg}}
\end{aligned}
\end{equation}
where $\ket{\Phi_0^k}$ are single-atom Rabi ground states with the $k$-atom,
whereas $\ket{\Psi_{0gg}}$ is the two-atoms Rabi ground state. It is  
apparent that STIRAP may occurr via the entangled intermediate state 
$\big[\ket{\Phi_0^1} \ket{u^2}+
\ket{\Phi_0^2} \ket{u^1} \big]/\sqrt{2}$ with rescaled peak Rabi frequencies. 
In particular the Stokes effective field strength is larger since 
$c_{02}(g) \to \sqrt{2} c_{02}(g)$. However the pump field may be almost resonant with other transitions, and Eq.~(\ref{eq:lambda-5}) contains stray linkages with a competing target state $\ket{\Psi_{0gg}}$ (magenta curve in Fig.~\ref{fig:stark}b), which may spoil population transfer to $\ket{2uu}$. 
The difference between the single and the two-atom case, emphasized in Fig.~\ref{fig:stark}b, is due to this effect. It turns out that both the nonvanishing $E_{0gg}- 2 E_0$ due to the interaction and to the dynamical Stark shifts in the multilevel structure detune the two-photon \enquote{ladder} $\ket{0uu} \to \ket{\Psi_{0gg}}$ (see inset in Fig.~\ref{fig:stark}b). We mention that 
the stray process can be fully suppressed by using a slight detuning $\delta_p=\delta_s$ of $\omega_p$ and $\omega_s$, which preserves the two-photon resonance for the $\ket{0uu} \to \ket{2uu}$ transition. 

\section{Conclusions}
We have proposed a dynamical detection method of USC based on two-pulse interference, which uses an ancillary level to probe the structure of the ground state. By driving adiabatically the system with a STIRAP protocol virtual photons in the grould state are converted to real ones. This channel is amplified by coherence, and ideally guarantees $100\%$ efficiency. 
However hardware may pose limitations: in semiconductors the detection of THz photons is prohibilitive, whereas 
in superconducting architectures it is not clear how to implement the additional level $\ket{u}$ used in the 
$\Lambda$ scheme implicit in all proposals 
or by Raman oscillations~\cite{ka:213-stassisavasta-prl-USCSEP,ka:214-huanglaw-pra-uscraman,ka:217-falci-fortphys-fqmt}. 
A way out in superconducting architectures is using a Vee-STIRAP~\cite{ka:218-falci-usc} technique, which may provide a unique dynamical tool for detecting virtual photons dressing excited states in available devices.
The STIRAP protocol is robust against imperfections of the drives being sensitive essentially to solid-state decoherence~\cite{kr:214-paladino-rmp} in the trapped subspace~\cite{ka:213-falci-prb-stirapcpb} which in our case is spanned by the $n=0,2$ Fock states of the cavity. Decoherence decreases the efficiency but this is not a severe problem since in the worst case it is larger than $30\%$~\cite{ka:213-falci-prb-stirapcpb}, and the very detection of photon pairs at the end of the protocol is already a \enquote{smoking gun} for USC.

\end{document}